\documentclass[pra,amsmath,amssymb,twocolumn,showpacs,aps,floatfix,superscriptaddress]{revtex4-1}
\usepackage{braket}
\usepackage{graphics}
\usepackage[usenames]{color}
\usepackage[dvips]{graphicx}
\usepackage{dcolumn}
\usepackage{natbib}
\usepackage{subfigure}
\usepackage{float}

\def\beq{\begin{equation}}
\def\eeq{\end{equation}}

\def \mU{\mathbb{U}}

\def \mN{\mathbb{N}}

\def \mC {\mathbb{C}}
\def \mS {\mathbb{S}}

\def \mV {\mathbb{V}}
\def \lt {\lambda t}

\newcommand{\la}{\langle}
\newcommand{\ra}{\rangle}
\newcommand{\lam}{\lambda}

\newcommand{\ga}{\gamma}
\newcommand{\Ga}{\Gamma}

\newcommand{\da}{\dagger}

\newcommand{\si}{\sigma}

\newcommand{\om}{\omega}

\newcommand{\de}{\delta}

\newcommand{\non}{\nonumber}
\newcommand{\pa}{\partial}

\def\prb#1{{ Phys.\ Rev. B\/} {\bf#1}}

\def\prl#1{{ Phys.\ Rev.\ Lett.} {\bf#1}}

\begin{document}
\title{Mutual Preservation of Entanglement}
\author{Andrzej Veitia}
\email{ ap3044@columbia.edu}
\affiliation{Optical Nanostructures Laboratory, Columbia University, New York, New York 10027, USA}
\author{Jun Jing}
\affiliation{Center for Controlled Quantum Systems and Department of Physics and Engineering Physics, Stevens Institute of Technology, Hoboken, New Jersey, 07030, USA}
\author{Ting Yu}
\affiliation{Center for Controlled Quantum Systems and Department of Physics and Engineering Physics, Stevens Institute of Technology, Hoboken, New Jersey, 07030, USA}
\author{Chee Wei Wong}
\affiliation{Optical Nanostructures Laboratory, Columbia University, New York, New York 10027, USA}
\begin{abstract}
 We study a generalized double Jaynes-Cummings (JC) model where two entangled pairs of two-level atoms interact indirectly.
 We focus on the case where the cavities and the entangled pairs are uncorrelated. We show that there exist initial states of
  the qubit system so that two entangled pairs are available at all times. In particular, the minimum entanglement in the pairs as a function of the
   initial state is studied. Finally, we extend our findings to a model consisting of multi-mode atom-cavity interactions. We use a non-Markovian quantum
   state diffusion (QSD) equation to obtain the steady-state density matrix for the qubits. We show that the multi-mode model also displays dynamical
   preservation of entanglement.
\end{abstract}
  \maketitle

\section{Introduction}
\begin{figure}[htb!]
\centering
\subfigure[\,Double JC model.  Two two-level atoms, prepared in entangled state $\rho_{A}$ interact, locally,  with uncorrelated single mode-cavities $F_{1}$ and $F_{2}$. ]
{
\includegraphics[scale=0.66]{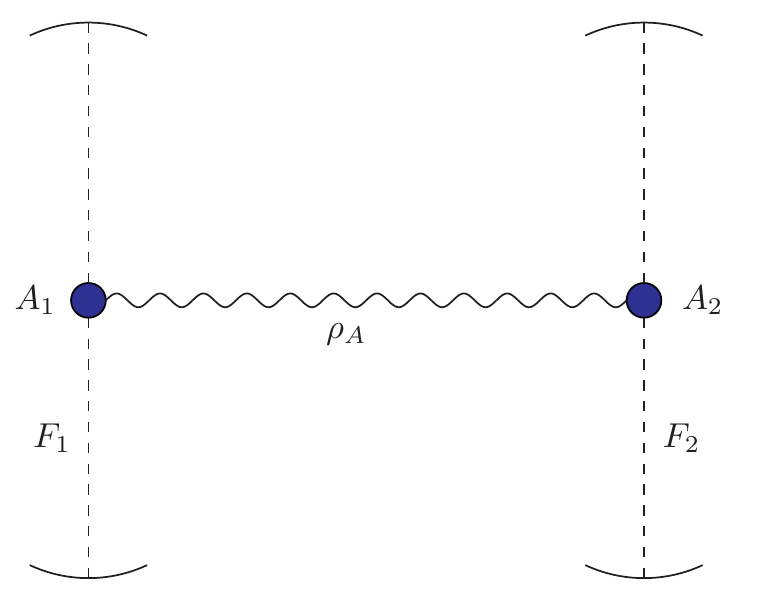}
\label{F:qubit-qubit}
}
\hspace{1.0 cm}
\subfigure[\,Generalized double JC model. The qubits $A_{1}$ and $B_{1(2)}$ interact indirectly via common cavity modes. We assume that, initially, entanglement is only present in the pairs $A_{1}A_{2}$ and $B_{1}B_{2}$.  ]
{
\includegraphics[scale=0.66]{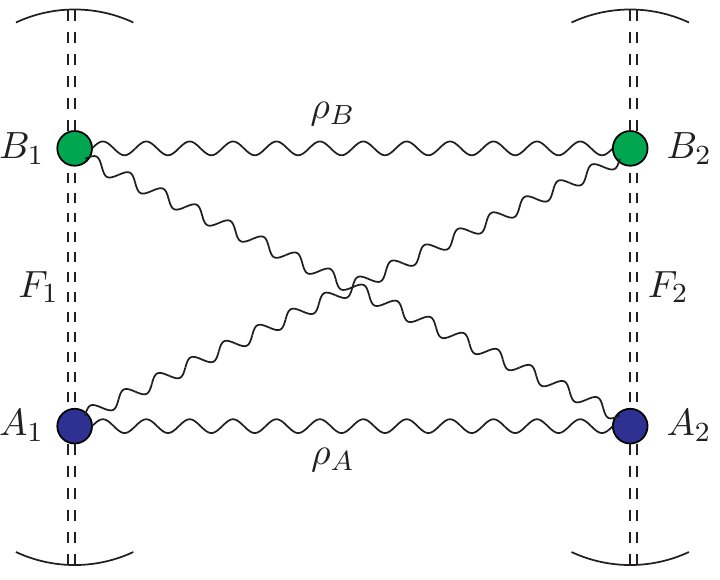}
\label{F:4qubits}
}
\caption[]{Double JC model (four-qubit model) and generalized double JC model (four-qubit-two-qutrit model). }
\end{figure}

The study of entanglement dynamics is crucial for the realization of quantum algorithms and quantum information processing protocols \cite{Chuang-Nielsen}. A significant number of works have been devoted to study the dynamics of quantum entanglement under environmental
 effects \cite{Yu-Eberly03,Yu-Eberly04,Jakobczyk-Jamroz04, Werlang09,Ficek-Tanas06,Yu-Eberly09}. Many previous studies focused on the simplest situation, namely, two-qubit entanglement dynamics. In reference \cite{Yu-Eberly04}, it was shown that contrary to what might be expected, the two-qubit entanglement can vanish completely in a finite time which is often referred to as ``entanglement sudden death" (ESD). One would naively expect the two-qubit entanglement to decay asymptotically as a result of noise-induced decoherence effects \cite{Yu-Eberly04}. ESD shows the fragility of entanglement under the unavoidable interaction with the environment. A simple model for ESD is that of two two-level atoms interacting via Jaynes-Cummings (JC) Hamiltonians with two uncorrelated single-mode cavities \cite{Yonac-Yu}.  This double JC model is schematically depicted in Fig.\ref{F:qubit-qubit}. Since the JC Hamiltonians conserve the number of excitations (atomic plus photonic), the model can be treated as a four-qubit network. It turns out that even when both cavities are prepared in the vacuum state the entanglement between the atoms dies and revives periodically. This behavior may be interpreted as periodic entanglement transfer between atomic and photonic systems \cite{Yonac-Yu,Yu-Eberly09}. This model has also been extended to the multi-mode case where the atom-cavity couplings are described by a spectral distribution. For this model it was shown that entanglement cannot be protected
regardless of the initial states (see \cite{Zhang09, Man10} and references therein). \\
\indent
The purpose of this paper is to study the preservation of entanglement in the network depicted in Fig.\ref{F:4qubits}. This model may be considered as an extension of the aforementioned double JC model. We assume that, initially, entanglement is only present in subsystems $A_{1}A_{2}$ and $B_{1}B_{2}$.  In this network, subsystems $A_{i}$ and $B_{i}$ ($i=1,2$) undergo excitation exchange interactions modeled via JC Hamiltonians.  We restrict our attention to the situation where the cavities are prepared in the vacuum state. Under these assumptions, the model may be considered a four-qubit (atoms) and two-qutrit (cavities) system. In this context, we show that the pairs $A_{1}A_{2}$ and $B_{1}B_{2}$ can be prepared in certain partially entangled states such that they remain entangled at all times. This is the main result of this work.\\
{\indent}The paper is organized as follows. In Sec.\ref{Yu-Eberly} we briefly discuss the double JC model and determine the corresponding evolution operator.  In Sec.\ref{single-mode} we examine the entanglement dynamics in more complex scenario as portrayed in Fig.(\ref{F:4qubits}). We derived a compact expression for the evolution operator corresponding to the case of single-mode qubit-cavity interaction. This facilitates the study of entanglement dynamics for different initial states of the system. We show that there exist initial states for this network, such that the entanglement between the distant parties ($A_{1}A_{2}$ and $B_{1}B_{2}$) never vanishes. Additionally,  we studied numerically the minimum entanglement in these pairs as a function of their initial state. In this sense we found the optimal initial states of the pairs which, surprisingly, do not turn out to be maximally entangled. We also study the emergence of entanglement in the initially separable pairs $A_{1}B_{2}$ and $A_{2}B_{1}$.\\
{\indent}Finally, in Sec.\ref{multi-mode} we include multimode atom-cavity interactions into our model. Here, the entanglement of the qubits is studied by means of a non-Markovian quantum state diffusion equations (QSD) \cite{Diosi,Gisin,Strunz}. In addition, the residual entanglement in the qubits is determined in the steady state limit of the QSD equation \cite{Zhao-Jing}. The analytical results obtained in section corroborate the numerical results reported in Sec.\ref{single-mode}.

\section{ENTANGLEMENT DYNAMICS IN THE DOUBLE JC MODEL}
\label{Yu-Eberly}
Let the Hamiltonian acting on system $(A_{i}F_{i})$ be $H^{(i)}=H_{0}^{(i)}+H_{int}^{(i)}$, where
 \begin{eqnarray}
 \label{double-JC1}
  H_{0}^{(i)}&=&\frac{\hbar}{2} \omega_{A_{i}}\sigma_{z}^{(A_{i})}+\hbar \omega_{i} {a_{i}}^{\dagger}a_{i},\\
  \label{double-JC2}
  H_{int}^{(i)}&=&\hbar{\lambda_{A_{i}}}(\sigma_{+}^{(A_{i})}a_{i}+\sigma_{-}^{(A_{i})}{a_{i}}^{\dagger})\\ \nonumber
 \end{eqnarray}
 and  $i=(1,2)$.
  The spectrum of this Hamiltonian is well known \cite{JC}. The knowledge of the energy eigenstates and eigenvectors could be used to determine the time evolution of the system. However, from a technical point of view, it is more convenient to find the time evolution operator by exponentiation of the Hamiltonian $H^{(i)}$, as described in \cite{Singh}. It turns that this method may be also applied to larger system as the one described in Fig.(\ref{F:4qubits}). For the double JC model we have:
 \beq
 \mU_{i}:=e^{-\frac{it}{\hbar}H^{(i)}}=e^{-i\omega t \hat{\mN}_{i}}e^{-i\lambda t \hat{\mC}_{i}}
\eeq
where $\hat{\mN}_{i}=a_{i}^{\dagger}a_{i}+\frac{1}{2}\sigma_{z}^{(A_{i})}$ and $\hat{\mC}_{i}=\sigma_{+}^{(A_{i})}a_{i}+\sigma_{-}^{(A_{i})}a_{i}^{\dagger}$. Here, we have assumed the zero detuning case ($\omega_{A_{i}}=\omega_{i}$) and used the relation $[\hat{\mN}_{i},\hat{\mC}_{i}]=0$. Now, one can easily show that
\beq
\label{U}
\mU_{i}=e^{-i\omega t \hat{\mN}_{i}}\left(
                        \begin{array}{cc}
                          \cos(\lambda t \sqrt{a_{i}a_{i}^{\dagger}}) & -i\frac{\sin(\lambda t \sqrt{a_{i}a_{i}^{\dagger}})}{\sqrt{a_{i}a_{i}^{\dagger}}}a_{i} \\
                          -i\frac{\sin(\lambda t \sqrt{a_{i}^{\dagger}a_{i}})}{\sqrt{a_{i}^{\dagger}a_{i}}}a_{i}^{\dagger} &  \cos(\lambda t\sqrt{a_{i}^{\dagger}a_{i}})\\
                        \end{array}
                      \right).
\eeq
Clearly, the time evolution operator for the joint system $A_{1}A_{2}F_{1}F_{2}$ is given  $\mU=\mU_{1}\otimes \mU_{2}$.  Following  \cite{Yonac-Yu,Yu-Eberly09}, we assume that both cavities are initially in the vacuum state while the atoms start out in one of the following partially entangled states:
\begin{eqnarray}
\label{Bell1}
\ket{\Phi_{A}}&=&\cos(\alpha)\ket{e_{A_{1}},e_{{A_{2}}}}+\sin(\alpha)\ket{g_{A_{1}},g_{A_{2}}}\\
\label{Bell2}
\ket{\Psi_{A}}&=&\cos(\alpha)\ket{e_{A_{1}},g_{{A_{2}}}}+\sin(\alpha)\ket{g_{A_{1}},e_{A_{2}}}.\\ \nonumber
\end{eqnarray}
Due to the fact that the JC Hamiltonian conserves the total number of excitations, the atomic reduced density matrix will be given by the X-state
\beq
\label{X}
\rho=\left(
       \begin{array}{cccc}
         a & 0& 0& f\\
         0&b& e& 0\\
         0& e^{*}& c&0 \\
         f^{*}&0&0&d \\
       \end{array}
     \right).
     \eeq

\indent
Throughout the present paper, we will quantify the entanglement $E(\rho)$ by means of  Wootter's concurrence  $C(\rho)$ \cite{Wotters}. For states of the form Eq.(\ref{X}), the concurrence can be written in the compact form
     \beq
     \label{conc}
   C(\rho)=2 \textrm{max}(0, |f|-\sqrt{bc}, |e|-\sqrt{ad}).
     \eeq
    Note that for the X-states of the form Eq.(\ref{Bell1}) and Eq.(\ref{Bell2}) we have $C(\rho)=|\sin(2\alpha)|$.\\

     \begin{figure}[htb!]
\centering
\subfigure[\,Concurrence as a function of time for system $A_{1}A_{2}$ when its initial state is given by  $\ket{\Phi_{A}}$. The curves correspond to   $\alpha=60^{\circ}$ (solid line), $\alpha=45^{\circ}$ (dashed line) and $\alpha=30^{\circ}$ (dotted line).  ]
{
\includegraphics[scale=0.8]{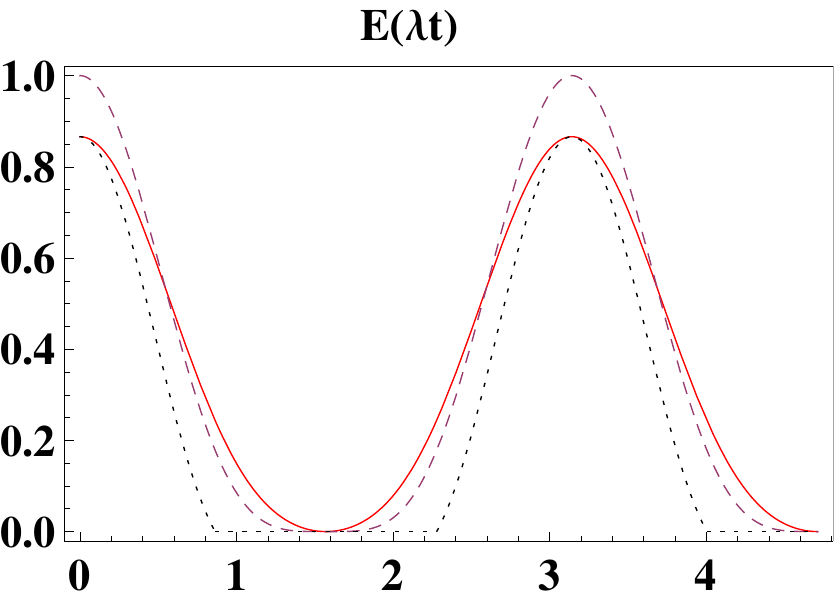}
\label{456085}
}
\subfigure[\,Concurrence as a function of time for system $A_{1}A_{2}$ when its initial state is given by  $\ket{\Psi_{A}}$             . The curves correspond to $\alpha=45^{\circ}$ (solid line), $\alpha=30^{\circ}$ (dashed line) and $\alpha=15^{\circ}$ (dotted line).]{
\includegraphics[scale=0.8]{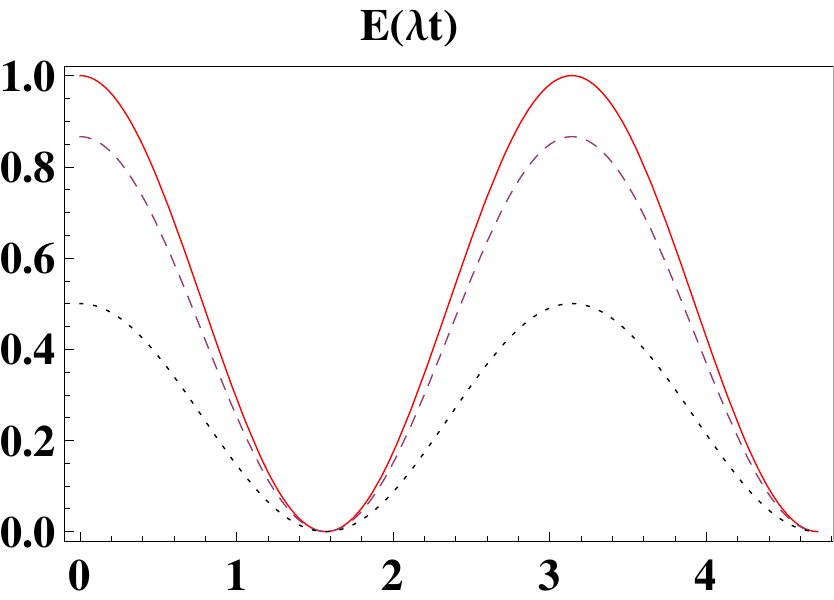}
\label{453015}
}
\caption[]{Evolution of entanglement in the double JC model \protect{ \cite{Yonac-Yu}}.}
\end{figure}

\indent
 Using Eq.(\ref{U}), one can determine the time-evolution of the reduced density matrix corresponding to the qubits $A_{1}A_{2}$ \cite{Yonac-Yu}. The entanglement dynamics is shown in Fig.\ref{456085} and Fig.\ref{453015}. Although both graphs describe the death and rebirth of entanglement, there are significant differences between Fig.\ref{456085} and Fig.\ref{453015}. In the case where the atoms start out in the state $\ket{\Phi_{A}}$ given by Eq.(\ref{Bell1}), the entanglement remains zero for finite periods of time (except when $\alpha=45^\circ$). These periods depend on the initial degree of entanglement in the system $A_{1}A_{2}$ (see Fig.\ref{456085}). On the other hand, when the atoms are prepared in the state $\ket{\Psi_{A}}$ given by Eq.(\ref{Bell2}), the entanglement decays to zero periodically and recovers immediately, independently of $\alpha$. Note that since we have assumed a symmetric scenario, the transformation $\alpha \rightarrow \pi/2-\alpha$ does not affect the concurrence.

\section{A MODEL OF MUTUAL PRESERVATION OF ENTANGLEMENT}
\label{single-mode}

\indent
In this section, we study the scenario depicted in Fig.(\ref{F:4qubits}). Here systems $A_{1},A_{2}, B_{1},B_{2}$ are assumed to be two-level atoms while $F_{1}$ and $F_{2}$ represent single-mode cavities.  Let the Hamiltonian acting on system $(A_{i}B_{i}F_{i})$ be $H^{(i)}=H_{0}^{(i)}+H_{int}^{(i)}$ where
 \begin{eqnarray}
 \label{ham0}
 H_{0}^{(i)}&=&\frac{\hbar}{2} \omega_{A_{i}}\sigma_{z}^{(A_{i})}+\frac{\hbar}{2}\omega_{B_{i}}\sigma_{z}^{(B_{i})}+\hbar \omega_{i} {a_{i}}^{\dagger}a_{i} , \\
  \label{ham-int}
  H_{int}^{(i)}&=&\hbar{\lambda_{A_{i}}}(\sigma_{+}^{(A_{i})}a_{i}+\sigma_{-}^{(A_{i})}{a_{i}}^{\dagger})+\hbar{\lambda_{B_{i}}}(\sigma_{+}^{(B_{i})}a_{i}+\sigma_{-}^{(B_{i})}
  {a_{i}}^{\dagger})\nonumber \\
 \end{eqnarray}
 and i=(1,2).
 The interaction of a single-mode quantized radiation field with
$N$ two-level atoms was first studied by Dicke \cite{Dicke54}. The spectrum corresponding to the Hamiltonian Eqs.(\ref{ham0})-(\ref{ham-int}) was found long ago \cite{Travis-Cummings68} and its associated dynamics has been extensively studied in \cite{Deng1985, Kumar-Mehta76,Smithers-Lu}. In addition, two-level atoms coupled to single-mode radiation field have been studied in connection with entanglement generation in cavity QED. \cite{Kim02, Plenio99}.\\

\indent
Following a similar method to that described in the previous section, we write the Hamiltonian $H^{(i)}$ as

  \beq
  H^{(i)}=\hbar \omega_{i}\hat{N}_{i}+\hbar \lambda_{A_{i}}\hat{C}^{(A_{i})}+\hbar \lambda_{B_{i}}\hat{C}^{(B_{i})}.
  \eeq
  with $\hat{N}_{i}={a_{i}}^{\dagger}a_{i}+\frac{1}{2}(\sigma_{z}^{(A_{i})}+\sigma_{z}^{(B_{i})})$,

  \begin{eqnarray}
   \hat{C}^{(A_{i})}&=&\epsilon_{A_{i}}\sigma_{z}^{(A_{i})}+\sigma_{+}^{(A_{i})}a_{i}+\sigma_{-}^{(A_{i})}{a_{i}}^{\dagger}\\
  \hat{C}^{(B_{i})}&=&\epsilon_{B_{i}}\sigma_{z}^{(B_{i})}+\sigma_{+}^{(B_{i})}a_{i}+\sigma_{-}^{(B_{i})}{a_{i}}^{\dagger},\\ \nonumber
  \end{eqnarray}
$\epsilon_{A_{i}}=\frac{\omega_{A_{i}}-\omega_{i}}{2\lambda_{A_{i}}}$ and $ \epsilon_{B_{i}}=\frac{\omega_{B_{i}}-\omega_{i}}{2\lambda_{B_{i}}}$.
  From now on, we shall assume that the atoms and cavities are identical, that is, $\lambda_{A_{1}}=\lambda_{A_{2}}=\lambda_{B_{1}}=\lambda_{B_{2}}=\lambda$ and $\omega_{1}=\omega_{2}=\omega$. In addition, we shall again restrict our attention to the zero detuning case i.e., $\epsilon_{A_{i}}=\epsilon_{B_{i}}=0$.  These assumptions, plus the fact that $[\hat{\mN}_{i}, \hat{C}^{(A_{i})}]=[\hat{\mN}_{i}, \hat{\mC}^{(B_{i})}]=0 $, allow us to write the local evolution operator as

\beq
 \mU_{i}=e^{-i H^{(i)} t}=e^{-i \omega t\hat{\mN}_{i}}e^{-i \lambda t \hat{\mC}_{i}}, \quad i=1,2\eeq
   where
\beq
\hat{\mC}_{i}:=\hat{\mC}^{(A_{i})}+\hat{\mC}^{(B_{i})}=\left(
  \begin{array}{cccc}
    0 & a_{i} & a_{i} & 0 \\
    a_{i}^{\dagger} & 0 & 0 & a_{i} \\
    a_{i}^{\dagger} & 0 & 0 & a_{i} \\
    0 & a_{i}^{\dagger} & a_{i}^{\dagger} & 0 \\
  \end{array}
\right)
\eeq
in the $\mathcal{H}_{A_{i}}\otimes \mathcal{H}_{B_{i}}$ basis given by $\ket{1^{(i)}}=\ket{e_{i},e_{i}}$,  $\ket{2^{(i)}}=\ket{e_{i},g_{i}}$, $\ket{3^{(i)}}=\ket{g_{i},e_{i}}$ and
$\ket{4^{(i)}}=\ket{g_{i},g_{i}}.$ The operator $\mU_{i}$ may be determined by exponentiating the matrix $\mC_{i}$. It can be shown that the even and odd powers of the operator $\mC_{i}$ read:

\beq
\hat{\mC}_{i}^{2k}=2^{k}\left(
  \begin{array}{cccc}
    a_{i} \mS_{i}^{k-1}a_{i}^{\dagger} & 0 & 0 & a_{i}\mS^{k-1}a_{i}\\
    0 & \mS^{k}/2 & {\mS^{k}}/2 & 0 \\
    0 & \mS^{k}/2 & {\mS^{k}}/2 & 0 \\
    a_{i}^{\dagger}\mS^{k-1}a_{i}^{\dagger}& 0 & 0 & a_{i}^{\dagger}\mS^{k-1}a_{i} \\
  \end{array}
\right), \quad k>0
\eeq

\beq
\hat{\mC}_{i}^{2k+1}=2^{k}\left(
             \begin{array}{cccc}
               0 & a_{i}\mS^{k} & a_{i}\mS^{k} & 0 \\
               \mS_{i}^{k}a_{i}^{\dagger} & 0 & 0 & \mS_{i}^{k}a_{i} \\
               \mS_{i}^{k}a_{i}^{\dagger} & 0 & 0 & \mS_{i}^{k}a_{i} \\
               0 & a_{i}^{\dagger}\mS_{i}^{k} & a_{i}^{\dagger}\mS_{i}^{k} & 0 \\
             \end{array}
           \right), \quad k\geq 0
        \eeq
where $\mS_{i}=a_{i}a_{i}^{\dagger}+a_{i}^{\dagger}a_{i}$. Writing $\mU_{i}=e^{-i\omega t \mN_{i}}\sum_{k=0}^{\infty} \frac{(-i \lambda t)^k}{k!}\hat{\mC}_{i}^{k}$, we obtain the following compact expression for the evolution operator $\mU_{i}$:
\begin{widetext}
\beq
\label{evolution}
\mU_{i}=e^{-i\omega t \hat{\mN}_{i}}\left(
          \begin{array}{cccc}
            1-2a_{i}{\mS_{i}}^{-1}\sin^{2}(\lambda_{i}t \sqrt{\frac{\mS_{i}}{2}}){a_{i}}^{\dagger} & -ia_{i}\frac{\sin(\lambda_{i}t \sqrt{2\mS_{i}})}{\sqrt{2\mS_{i}}} & -ia_{i}\frac{\sin(\lambda_{i}t \sqrt{2\mS_{i}})}{\sqrt{2\mS_{i}}} & -2a_{i}{\mS_{i}}^{-1}\sin^{2}(\lambda_{i}t \sqrt{\frac{\mS_{i}}{2}})a_{i} \\
            -i\frac{\sin(\lambda_{i}t \sqrt{2\mS_{i}})}{\sqrt{2\mS_{i}}}{a_{i}}^{\dagger} & \cos^{2}(\lambda_{i}t \sqrt{\frac{\mS_{i}}{2}}) & -\sin^{2}(\lambda_{i}t \sqrt{\frac{\mS_{i}}{2}}) & -i\frac{\sin(\lambda_{i}t \sqrt{2\mS_{i}})}{\sqrt{2\mS_{i}}}a_{i} \\
            -i\frac{\sin(\lambda_{i}t \sqrt{2\mS_{i}})}{\sqrt{2\mS_{i}}}{a_{i}}^{\dagger}& -\sin^{2}(\lambda_{i}t \sqrt{\frac{\mS_{i}}{2}}) & \cos^{2}(\lambda_{i}t \sqrt{\frac{\mS_{i}}{2}}) &  -i\frac{\sin(\lambda_{i}t \sqrt{2\mS_{i}})}{\sqrt{2\mS_{i}}}a_{i} \\
              -2{a_{i}}^{\dagger}{\mS_{i}}^{-1}\sin^{2}(\lambda_{i}t \sqrt{\frac{\mS_{i}}{2}}){a_{i}}^{\dagger}& -i{a_{i}}^{\dagger} \frac{\sin(\lambda_{i}t \sqrt{2\mS_{i}})}{\sqrt{2\mS_{i}}}&-i{a_{i}}^{\dagger} \frac{\sin(\lambda_{i}t \sqrt{2\mS_{i}})}{\sqrt{2\mS_{i}}}  & 1-2{a_{i}}^{\dagger}{\mS_{i}}^{-1}\sin^{2}(\lambda_{i}t \sqrt{\frac{\mS_{i}}{2}}){a_{i}} \\
          \end{array}
        \right).
\eeq
 \end{widetext}
Clearly, the time evolution for the joint system  $A_{1}A_{2}B_{1}B_{2}F_{1}F_{2}$ is given by $\mU_{1}\otimes \mU_{2}$.
  We consider the situation where systems $A_{1}A_{2}$ and $B_{1}B_{2}$ (systems A and B) are initially prepared in entangled pure states $\rho_{A}=\ket{\phi_{A}}\bra{\phi_{A}}$ and $\rho_{B}=\ket{\phi_{B}}\bra{\phi_{B}}$.  In addition, we assume that there are no additional correlations present in the total system. Thus, the initial density operator may be written as  $\rho_{0}=\ket{\phi_{A}}\bra{\phi_{A}}\otimes \ket{\phi_{B}}\bra{\phi_{B}} \otimes \rho_{F_{1}} \otimes \rho_{F_{2}}$. At later times we have:
   \beq
   \label{evol}
   \rho=\mU_{1}\otimes\mU_{2} \rho_{0} {\mU_{1}}^{\dagger}\otimes {\mU_{2}}^{\dagger}.
   \eeq

  Following the double JC model \cite{Yonac-Yu,Yu-Eberly09} discussed in the previous section,  we assume that the states $\ket{\phi_{A}}$ and $\ket{\phi_{B}}$ are of the form

 \begin{eqnarray}	
 \label{Phi}
 \ket{\Phi_{A(B)}}&=&\cos(\alpha)\ket{e_{A_{1}(B_{1})},e_{A_{2}(B_{2})}} \nonumber \\
 &+&\sin(\alpha)\ket{g_{A_{1}(B_{1})},g_{{A_{2}}(B_{2})}}
 \end{eqnarray}
 or
 \begin{eqnarray}
 \label{Psi}
 \ket{\Psi_{A(B)}}&=&\cos(\alpha)\ket{e_{A_{1}(B_{1})},g_{{A_{2}}(B_{2})}} \nonumber \\
 &+&\sin(\alpha)\ket{g_{A_{1}(B_{1})},e_{A_{2}(B_{2})}}.
 \end{eqnarray}
 In either case, we may write $\ket{\phi_{A}}=\sum_{k} s_{k}\ket{\phi_{A_{1},k},\phi_{A_{2},k}}$. Making use of equations Eq.(\ref{evolution}), Eq.(\ref{evol}) and tracing out the degrees of freedom of systems $B_{1}B_{2}$ and $F_{1}F_{2}$, we obtain the following expression for the reduced density matrix corresponding to qubits $A_{1}$ and $A_{2}$ (system A):
\begin{equation}
\label{main1}
\rho^{A}_{kl,m n}=\sum_{i,j} s_{i}s_{j}\textrm{Tr}_{B}(\rho_{B} \mV^{(B_{1})}_{i m k j}\otimes \mV^{(B_{2})}_{\pi(i) n l \pi(j)}).
\end{equation}
 Here $\pi(1)=1,\quad  \pi(2)=2$, for partially entangled states of the form Eq.(\ref{Phi}) while $\pi(1)=2, \pi(2)=1$, for partially entangled states of the form Eq.(\ref{Psi}). The above $\mV^{B_{1(2)}}_{ijkl}$ operators are computed from the evolution operator Eq.(\ref{evolution}). They are given by
\beq
\label{main2}
\mV^{(B_{1(2)})}_{ijkl}=\textrm{Tr}_{F_{1(2)}}(\rho_{F_{1(2)}}\braket{i|\mU_{1(2)}^{\dagger}|j}_{A_{1(2)}}\cdot\braket{k|\mU_{1(2)}|l}_{A_{1(2)}})
\eeq
where $\ket{1}=\ket{e}$ and $\ket{2}=\ket{g}$.  In the appendix Sec.\ref{appendix}, we list the set of non-vanishing operators $\mV^{(B_{1(2)})}_{ijkl}$ for the case where the cavities have a well defined number of excitations (i.e. $ \rho_{F_{i}}=\ket{N}\bra{N} $). Note that expression Eq.(\ref{main1}) also holds true in the case where the systems $A_{1}A_{2}$ and $B_{1}B_{2}$ are prepared in different types of states, e.g., $\ket{\Phi_{A}}\otimes \ket{\Psi_{B}}$. From symmetry considerations, we easily see that if the qubits start out in either $\ket{\Phi_{A}}\otimes \ket{\Phi_{B}}$ or $\ket{\Psi_{A}}\otimes \ket{\Psi_{B}}$, then $\rho^{A}=\rho^{B}$ at all times.

Similarly, we can write down an expression for the reduced density matrix for the qubit pairs $A_{1} B_{2}$ and $A_{2}B_{1}$. For simplicity, we shall consider only the situation when the initial state of the qubits is of the form $\ket{\Phi_{A}}\otimes\ket{\Phi_{B}}$ or $\ket{\Psi_{A}}\otimes\ket{\Psi_{B}}$. Then we obtain:
\beq
\label{main3}
\rho^{A_{1}B_{2}}_{kl,mn}= \sum_{i,j,p,q} s_{i}\bra{i}V_{\pi(j)nl\pi(p)}\ket{q}s_{q} s_{j}\bra{j}V_{\pi(i)mk\pi(q)}\ket{p} s_{p}.
\eeq
 The usefulness of expressions Eq.(\ref{main1}), Eq.(\ref{main2}) and Eq.(\ref{main3}) lies in the fact that they can be evaluated in an automated fashion. They can also be applied to the more general case in which the cavities are prepared in mixed states \cite{Veitia}.

  \subsection{Partially entangled Bell States $\ket{\Phi_{A}}$ and $\ket{\Phi_{B}}$.}

 We start by considering the case where systems ${A_{1}A_{2}}$ and ${B_{1}B_{2}}$ are both initially in the same partially entangled state of the form Eq.(\ref{Phi}). As mentioned before, in the symmetric scenario in which the cavities are initially in the same quantum state, it suffices to compute the reduced density matrix of one of the systems, say $A_{1}A_{2}$. Making use of Eq.(\ref{main1}) we determine the non-vanishing matrix elements
\begin{eqnarray}
\rho^{A}_{11}&=&a^2 \cos^{4}(\alpha)+\frac{b^2+h^2+2p^{2}}{4}\sin^{2}(2 \alpha)\nonumber \\
&+& k^{2}\sin^{4}(\alpha)\\
\label{221}
\rho^{A}_{22}&=& a d\cos^{4}(\alpha)+ \frac{b f+h m-2p^{2}}{4}\sin^{2}(2\alpha)\nonumber \\
&+&k n\sin^{4}(\alpha)\\
\rho^{A}_{33}&=&\rho^{A}_{22}\\
\rho^{A}_{44}&=&d^{2}\cos^{4}(\alpha)+\frac{f^{2}+2p^{2}+m^{2}}{4}\sin^{2}(2\alpha)\nonumber \\
&+& n^{2}\sin^{4}(\alpha)\\
\label{141}
\rho^{A}_{14}&=&{\rho^{A}_{41}}^{*}=\frac{1}{2}e^{-2i\omega t}((c^{2}+q^{2})\cos^{2}(\alpha)  \nonumber \\
&+& (l^{2}+r^{2})\sin^{2}(\alpha))\sin(2\alpha).\\ \nonumber
\end{eqnarray}
 The functions $a,b,c \ldots$ can be found in the appendix Sec.\ref{appendix}. Of particular interest is the situation where both cavities are initially in the ground state, that is $\rho_{F_{i}}=\ket{0_{i}}\bra{0_{i}}$ for $\quad i=(1,2)$. \\

\begin{figure}[htb!]
\centering
\subfigure[\,Concurrence as a function of time for the cases  $\alpha=45^{\circ}$ (solid line), $\alpha=60^{\circ}$ (dashed line) and $\alpha=85^{\circ}$ (dotted line).]
{
\includegraphics[scale=0.8]{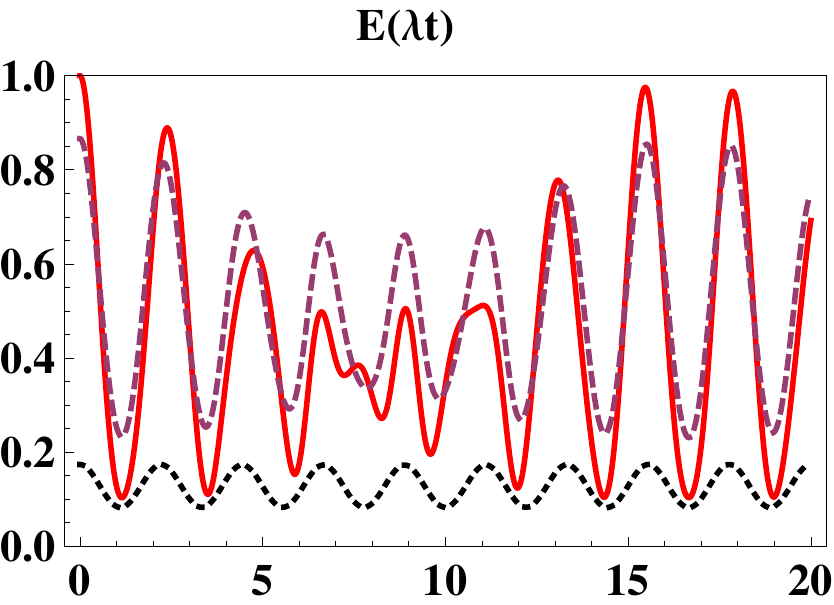}
\label{phi-phi45-60-85}
}

\subfigure[\,Concurrence as a function of time for the cases $\alpha=37^{\circ}$ (solid line) and $\alpha=15^{\circ}$ (dashed line).]{
\includegraphics[scale=0.8]{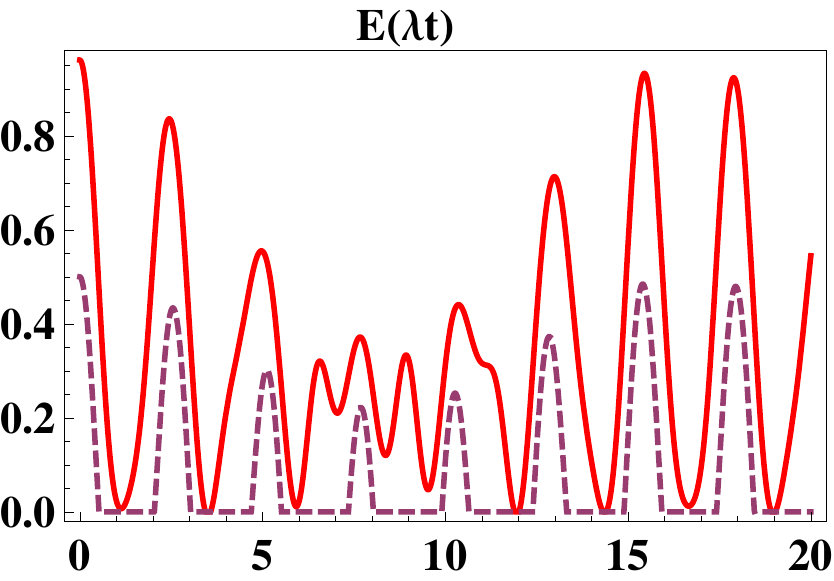}
\label{phi-phi37-15}
}
\caption[]{Evolution of entanglement for systems $A_{1}A_{2}$ ($ B_{1}B_{2}$). The qubits are initially prepared in  $\ket{\Phi_{A}}\otimes \ket{\Phi_{B}}.$}
\end{figure}

\begin{figure}[htb!]{
\includegraphics[scale=0.8]{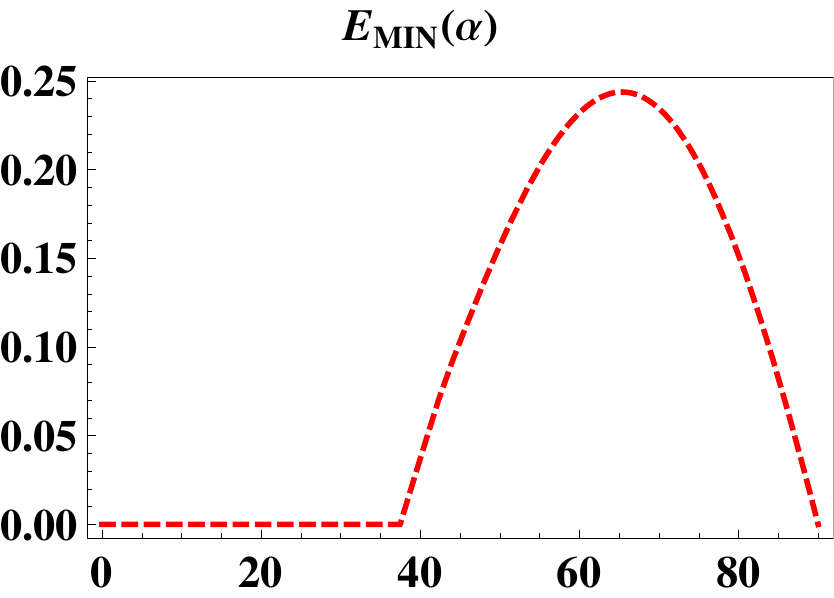}
\caption[]{Minimum Entanglement in $A_{1}A_{2}$ ($B_{1}B_{2}$) as a function of $\alpha$. }
\label{minimum-ent}}
\end{figure}

\begin{figure}[htb]
\centering
\subfigure[\,Concurrence as a function of time for $A_{1}B_{2}$ (dashed line) and $A_{1}A_{2}$ (solid lines). Here $\alpha=75^\circ$.]
{
\includegraphics[scale=0.8]{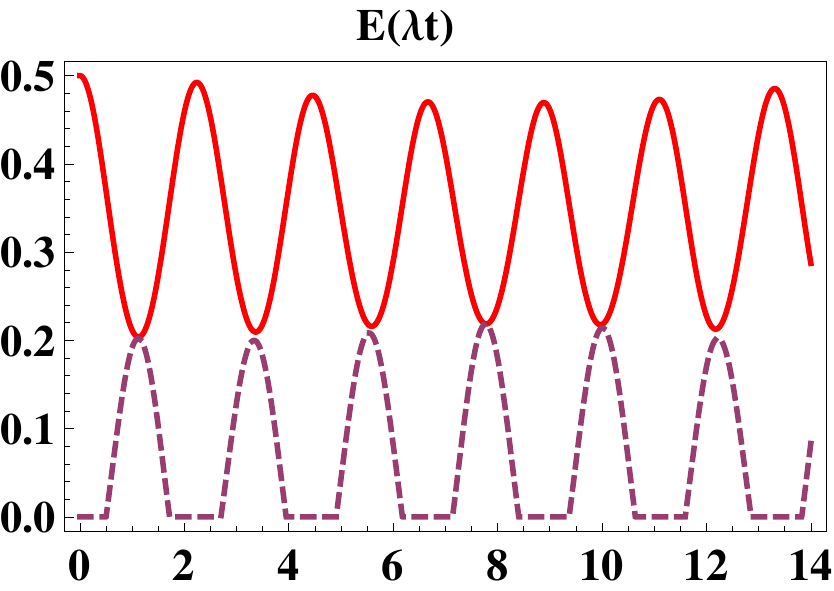}
\label{diag-75}
}
\hspace{1.0 cm}
\subfigure[\,Concurrence as a function of time for $A_{1}B_{2}$ (dashed line) and $A_{1}A_{2}$ (solid lines). Here $\alpha=45^\circ$.]
{
\includegraphics[scale=0.8]{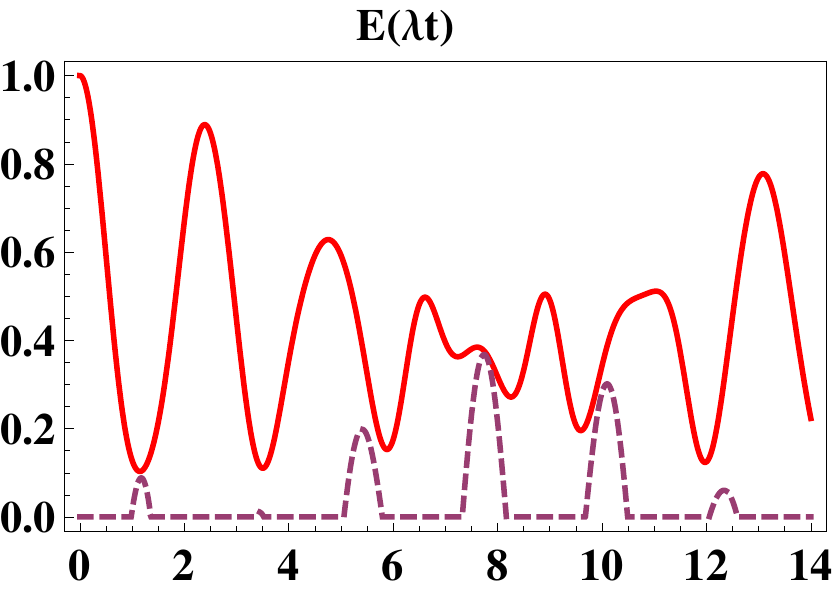}
\label{diag-45}
}
\caption[]{Evolution of entanglement for systems  $A_{1}A_{2}$ ($B_{1}B_{2}$) (solid line) and $A_{1}B_{2}$ ( $A_{2}B_{1}$) (dashed line). The qubits are initially prepared in  $\ket{\Phi_{A}}\otimes \ket{\Phi_{B}}.$}
\end{figure}

\indent
It turns out that for certain values of $\alpha$, the pairs $A_{1}A_{2}$ and $B_{1}B_{2}$ remain entangled at all times (see Fig.(\ref{phi-phi45-60-85})). It is interesting to study the minimum entanglement $E_{min}=\min_{t}C(\rho^{A}(t))$ in these pairs as a function of $\alpha$. Based on numerical analysis, we conclude that for $ 37.2^{\circ} < \alpha < 90^{\circ}$, there is always some residual entanglement in systems $A_{1}A_{2}$ and $B_{1}B_{2}$, as shown in Fig.(\ref{minimum-ent}). This result is corroborated by Fig.(\ref{phi-phi37-15}) where the time evolution of entanglement is shown for some values of $\alpha <37^{\circ}$.  If we adopt $E_{min}$ as a measure of the robustness of entanglement, we see from Fig.(\ref{phi-phi37-15}) that the most resilient state corresponds to $\alpha \approx 65.5^{\circ}$ for which $C(\rho^{A})>0.24$. Interestingly, it does not correspond to the maximally entangled state ($\alpha=45^{\circ}$). This non-trivial reflects a trade-off between the initial energy of the system and its entanglement. As $\alpha$ approaches $90^{\circ}$, the initial state of system approaches the energy eigenstate $\ket{g,g}\otimes\ket{g,g}\otimes\ket{0_{1}}\otimes\ket{0_{2}}$. Consequently, its small amount of entanglement will not vary considerably with time. On the other hand, when $\alpha< 45^{\circ}$, the pairs ($A_{1}A_{2}$ and $B_{1}B_{2}$) are more likely to be excited which renders dynamics of system more complex and tends to degrade the entanglement in the pairs. Moreover, the initial entanglement goes to zero as $\alpha$ approaches zero degrees. These two facts combined give rise to the critical value $\alpha_{cr} \approx 37$ such that the entanglement in the pairs vanishes for finite periods of time when $\alpha< \alpha_{cr}$.\\
\indent It is also important to mention that in order to retain some entanglement in the pairs $A_{1}A_{2}$ and $B_{1}B_{2}$, they must both be initially entangled.
   One can show that if the qubits start out in the state $\ket{\Phi_{A}}\otimes \ket{g}_{B_{1}}\otimes \ket{g}_{B_{2}}$, then entanglement in $A_{1}A_{2}$ will vanish for finite periods of time. Some entanglement will be transferred to $B_{1}B_{2}$ and for certain values of $\alpha$ it is possible to have one entangled pair at all times \cite{Veitia}.
\\
{\indent}It is also interesting to look at the entanglement between qubits $A_{1}$ and $B_{2}$ ($A_{2}$ and $B_{1}$). Here we shall also consider a symmetric configuration. Thus,
it suffices to determine the density matrix for one of pairs, say $A_{1}B_{2}$. Making use of Eq.(\ref{main3}) we obtain
\begin{eqnarray}
\rho_{11}^{A_{1}B_{2}}&=&a^{2}\cos^{4}(\alpha)+\frac{bh+p^2}{2}\sin^{2}(2\alpha) \nonumber \\
&+& k^{2}\sin^{4}(\alpha)\\
\label{222}
\rho_{22}^{A_{1}B_{2}}&=& a d \cos^{4}(\alpha)+\frac{f h+b m -2 p^{2}}{4}\sin^{2}(2\alpha)\nonumber \\
&+& k n \sin^{4}(\alpha)\\
\rho_{33}^{A_{1}B_{2}}&=&\rho_{22}^{A_{1}B_{2}}\\
\rho_{44}^{A_{1}B_{2}}&=& d^{2}\cos^{4}(\alpha)+\frac{f m+ p^{2}}{2}\sin^{2}(2\alpha)\nonumber \\
&+& n^{2}\sin^{4}(\alpha)\\
\label{142}
\rho_{14}^{A_{1}B_{2}}&=&{\rho^{A_{1}B_{2}}_{41}}^{*}=e^{-2i \omega t}(c q \cos^{2}(\alpha)\nonumber \\
&+& l r \sin^{2}(\alpha))\sin(2\alpha),\\ \nonumber
\end{eqnarray}
where the functions $a,b,c \ldots$ can be found in the appendix Sec.\ref{appendix}.\\
Note that the pairs $A_{1}B_{2}$ and $A_{2}B_{1}$ start out in separable (mixed) states. As a result of indirect interactions between $A_{i}B_{i}$, some fraction of the original entanglement in $A_{1}A_{2}$ and $B_{1}B_{2}$ will be transferred to these pairs. We assume that the cavities are prepared in the vacuum state. The case where $A_{1}A_{2}$ and $B_{1}B_{2}$ are initially in the state $\ket{\Phi_{A}}\otimes \ket{\Phi_{B}}$ with $\alpha=75^{\circ}$ is shown in Fig.(\ref{diag-75}). From this graph we see that for certain periods of time, we have the choice of selecting either two entangled or two separable pairs. Note that for this value of $\alpha$, the concurrences exhibit an approximately sinusoidal behavior. The dynamics corresponding to $\alpha=45^{\circ}$  turns out to be far more complex as shown in Fig.(\ref{diag-45}).
  Note that these graphs suggest that the entanglement in $A_{1}B_{2}$ can never exceed that in $B_{1}B_{2}$ (or $A_{1}A_{2}$) . In fact, this is a direct consequence of equations Eqs.(\ref{221}), (\ref{141}), (\ref{222}) and (\ref{142}). Using the inequalities $cq \leq \frac{1}{2}(c^{2}+q^{2})$ and $lr \leq \frac{l^{2}+r^{2}}{2}$ one proves that $|\rho_{14}^{A_{1}B_{2}}|\leq |\rho_{14}^{A}|$. In addition, we have
  \begin{eqnarray}
  \rho_{22}^{A}-\rho_{22}^{A_{1}B_{2}}&=&\frac{(h-b)(m-f)}{4}\sin^{2}(2\alpha)\nonumber \\
  &=&-\frac{1}{4}\cos^{2}(2\lambda t \sqrt{N+\frac{1}{2}})\sin^{2}(2\alpha)\leq 0\nonumber \\
  \end{eqnarray}
  which completes the proof.

\subsection{Partially entangled Bell States $\ket{\Psi_{A}}$ and $\ket{\Psi_{B}}$}
It turns out that if systems $A_{1}A_{2}$ and $B_{1}B_{2}$ are initially in states of the form in Eq.(\ref{Psi}), their entanglement cannot be preserved, for any value of $\alpha$. Moreover, as in the previous subsection, the concurrence of $A_{1}B_{2}$ ($A_{2}B_{1}$) never exceeds that of $A_{1}A_{2}$ ($B_{1}B_{2}$).
 Using expression Eqs.(\ref{main1}) and (\ref{main2}),  we compute the density matrix describing $A_{1}A_{2}$ and $B_{1}B_{2}$. The nonvanishing matrix elements read

\begin{eqnarray}
\rho^{A}_{11}&=& a k \cos^{4}(\alpha)+ \frac{ b h+p^{2}}{2}\sin^{2}(2 \alpha)+a k\sin^{4}(\alpha)\\
\rho^{A}_{22}&=& a n \cos^{4}(\alpha)+\frac{f h +b m -2 p^{2}}{4}\sin^{2}(2\alpha) \nonumber \\
&+& d k \sin^{4}(\alpha)\\
\rho^{A}_{33}&=& d k \cos^{4}(\alpha)+\frac{f h +b m -2 p^{2}}{4}\sin^{2}(2\alpha)\nonumber \\
&+& a n \sin^{4}(\alpha)\\
\rho^{A}_{44}&=&d n \cos^{4}(\alpha)+\frac{f m +p^{2}}{2}\sin^{2}(2\alpha)\nonumber \\
&+& d n \sin^{4}(\alpha)\\
\rho^{A}_{23}&=&{\rho^{A}_{32}}^{*}=\frac{c l +q r}{2}\sin(2\alpha).\\ \nonumber
\end{eqnarray}

\begin{figure}[htb]
\centering
\subfigure[\, Concurrence as a function of time for the cases $\alpha=25^{\circ}$ (solid line), $\alpha=45^{\circ}$ (dashed line) and $\alpha=60^{\circ}$ (dotted line).]
{
\includegraphics[scale=0.8]{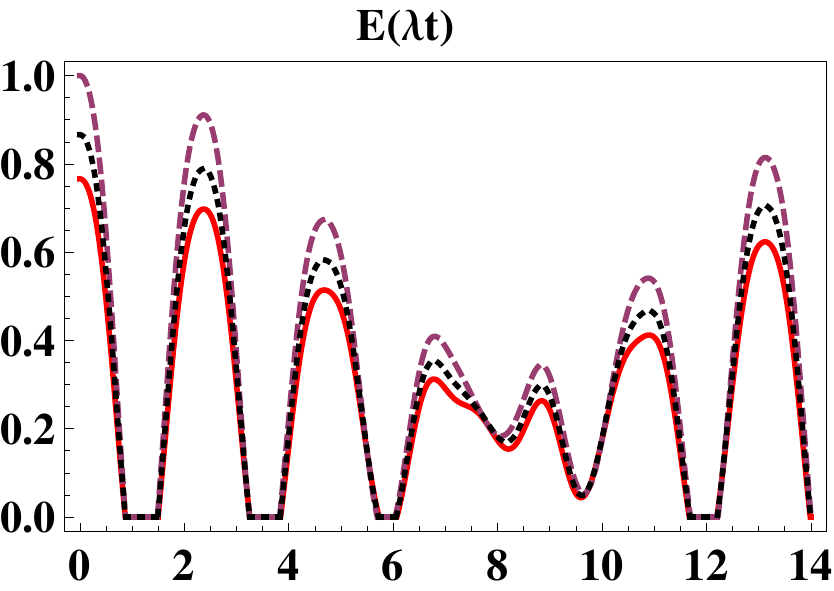}
\label{psi-psi25-45-60}
}
\hspace{1.0 cm}
\subfigure[ \, Concurrence as a function  of time (zoomed in) for the cases  $\alpha=25^{\circ}$ (solid line) and $\alpha=45^{\circ}$ (dashed line). ]
{
\includegraphics[scale=0.8]{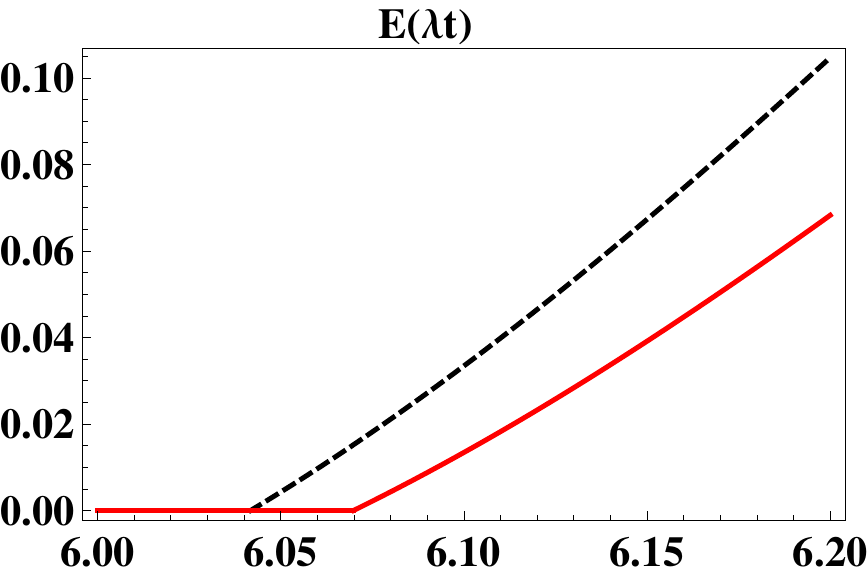}
\label{zoom}
}
\caption[]{Evolution of entanglement for systems $A_{1}A_{2}$ ($ B_{1}B_{2}$). The qubits are initially prepared in  $\ket{\Psi_{A}}\otimes \ket{\Psi_{B}}.$}
\end{figure}

 The curves corresponding to entanglement as a function of time for different values of $\alpha$ are shown in Fig.(\ref{psi-psi25-45-60}). We found that the concurrence vanishes for finite periods of time for every $\alpha$. Note that the curves appear to coalesce and go to zero simultaneously  regardless of the value $\alpha$. However if we zoom in on a portion of the graph, we can see that this is not the case (see Fig.(\ref{zoom})).
  We conclude this section with the following remark. Expressions Eqs.(\ref{main1}) and (\ref{main2}) may also be applied to the situation when the cavity modes are excited, i.e. $N>1$. We found that entanglement cannot be preserved (in the sense of Fig.\ref{phi-phi45-60-85}) unless both cavities are prepared in the vacuum state ($N=0$). The same holds true for the $\ket{\Psi_{A}} \otimes \ket{\Psi_{B}}$ case.

\section{Multimode Interaction}
\label{multi-mode}
 In this section we extend our analysis to the case where the two-level atoms interact with multi-mode cavities (structured environment). We describe this situation with the following natural generalization of the Hamiltonian Eq.(\ref{ham0}) and Eq.(\ref{ham-int}):
 \begin{eqnarray}
 H_{0}^{(i)}&=& H_{AB}^{(i)}+H_{F}^{(i)}\\
 H_{AB}^{(i)}&=&\frac{\hbar\omega_{i}}{2}(\sigma_{z}^{(A_{i})}+\sigma_{z}^{(B_{i})})\\
H_{F}^{(i)}&=&\sum_{k}\hbar \omega_{ik}{a_{ik}}^{\dagger}a_{ik}\\
H_{int}^{(i)}&=&\sum_{k}\hbar\lambda_{ik}^*(\sigma_{i}^{(A_{i})}+\sigma_{-}^{(B_{i})})a_{ik}^{\dagger}+h.c.
 \end{eqnarray}
 Here, $a_{ik}$ and $a_{ik}^{\dagger}$, correspond to the creation and annihilation operators of the $k^{th}$ electromagnetic mode in the $i^{th}$ cavity and frequency $\omega_{ik}$.
Using the Bargmann state of the baths to trace out the baths' degree of
freedom, we have
\begin{eqnarray}\nonumber
i\pa_t\psi_t(z^*)&=&\sum_{i=1}^2(H_{AB}^{(i)}
+L_i\sum_k \lambda_{i k}^*z_{i k}^*e^{i\om_{i k}t}\\
&+&L_i^\dagger \sum_{k} \lambda_{i k}e^{-i\om_{i k}t}\frac{\pa}{\pa z_{i k}^*})\psi_t(z^*)
\end{eqnarray}
where $\psi_t(z^*)$ is the system stochastic vector for the four qubit system $A_{1}A_{2}B_{1}B_{2}$ and
$L_i \equiv\si_-^{(A_{i})}+\si_-^{(B_{i})}$. The reduced density matrix is constructed from
\beq
\rho=M[|\psi_t(z^*)\ra\la\psi_t(z^*)|]=\int\frac{d^{2}z^2}{\pi}e^{-|z|^{2}}\ket{\psi_{t}(z^{*})}\bra{\psi_{t}(z^{*})}.
\eeq

Since there is no direct interaction between the subsystems $(A_1,B_1, F_1)$ and $(A_2,B_2,F_2)$, the
noises generated by the two local baths and O-operators are uncorrelated. The zero temperature assumption together with the chain rule
$\frac{\pa}{\pa z_{ i k}^*}=\int_0^tds\frac{\pa z_{ i s}^*}{\pa
z_{i k}^*}\frac{\de}{\de z_{i s}^*}$, allow us to construct the following QSD equation
\begin{eqnarray}
\pa_t\psi_t(z^*)&=&\sum_{i=1}^2 [-iH_{AB}^{(i)}
+L_i z_{i t}^* \non \\
&-& L_i^\da\int_0^tdsG_j(t,s)O_i(t,s,z_i^*)]\psi_t(z^*), \non \\
&\equiv&\sum_{j=1}^2\left[-iH_{AB}^{(i)}
+L_i z_{i t}^*-L_i^\da\bar{O}_i(t,z_i^*)\right]\psi_t(z^*),\nonumber \\
\end{eqnarray}
where $z_{i t}^*=-i\sum_\lam \lambda_{i k}^*z_{i k}^*e^{i\om_{i k}t}$,
$G_i(t,s)=\sum_k|\lambda_{i k}|^2e^{-i\om_{i k}(t-s)}$ and
$O_i(t,s,z_i^*)\psi_t(z^*)=\frac{\de}{\de z_{is}^*}\psi_t(z^*)$.\\
The QSD method yields $\bar{O}_i(t,z^*)=F_{i1}(t)O_1+F_{i2}(t)O_2
+iU_i(t,z^*_i)O_3$, where $O_1=L$,
$O_2=\sigma^{(A)}_z\sigma^{(B)}_-+\sigma^{(A)}_-\sigma^{(B)}_z$,
$O_3=\sigma^{(A)}_-\sigma^{(B)}_-$, and $U_i(t,z^*_i)\equiv\int_0^tdsU_i(t,s)z_{is}^*$
to be determined. Assuming
$G_i(t,s)=\frac{\Ga\ga_i}{2}e^{-\ga_i|t-s|}$, which corresponds to the Lorentz spectral density for the multi-mode cavities
$S(\om_j)=\frac{1}{2\pi}\frac{\Ga\ga_j^2}{\om_j^2+\ga_j^2}$, we obtain
\begin{eqnarray}
\partial_tF_{i1}(t)&=&\frac{\Ga\ga_i}{2}+(-\ga_i+i\om_i)F_{i1}+F_{i1}^2
+3F_{i2}^2\non \\
&-&\frac{i}{2}\bar{U}_i, \\
\partial_tF_{i2}(t)&=&(-\ga_i+i\om_i)F_{i2}-F_{i1}^2+4F_{i1}F_{i2}
+F_{i2}^2\non \\
&-&\frac{i}{2}\bar{U}_i, \\
\label{FFU} \partial_t\bar{U}_i(t)&=&-2i\ga_i
F_{i2}+(-2\ga_i+2i\om_i)\bar{U}_i \non \\
&+& 4F_{i1}\bar{U}_i,
\end{eqnarray}
where $\bar{U}_i(t)\equiv\int_0^tdsG_i(t,s)U_i(t,s)$. The boundary conditions
are given by $F_{i1}(0)=F_{i2}(0)=\bar{U}_i(0)=0$ and $U_i(t,t)=-4iF_{i2}(t)$.

It is also known that  an open system in a non-Markovian bath can approach a stable final state in the long time limit, as long as the bath correlation function has a well-defined Markov limit. For our model we have that  $\gamma_i\rightarrow\infty$ implies $G(t,s)\rightarrow\Gamma\delta(t,s)$. For notational simplicity,  we write the density matrix for the total system as
\begin{equation}
\rho=\left(\begin{array}{cccc}
      a & b & c & d \\
      e & f & g & h \\
      i & j & k & l \\
      m & n & o & p\\
    \end{array}\right)
\end{equation}
where $a,b,d,\cdots,p$ represent a $4\times4$ sub-matrices. We work in the basis   \{$\ket{e,e,e,e}, \ket{e,g,e,e}\ldots \ket{g,g,g,g} \}\in \mathcal{H}_{A_{1}}\otimes\mathcal{H}_{B_{1}}\otimes\mathcal{H}_{A_{2}}\otimes\mathcal{H}_{B_{2}} $. The final stable state for
this model is found to be
\begin{equation}
\label{long-time}
\rho_\infty=\left(\begin{array}{cccc}
      0 & 0 & 0 & 0 \\
      0 & F & G & H \\
      0 & J & K & L \\
      0 & N & O & P\\
    \end{array}\right),
\end{equation}
where
\begin{equation}
F=K=-G=-J, \quad H=-L=N^\da=-O^\da,
\end{equation}
and each non-vanish sub-matrix has the form
\begin{equation}
\left(\begin{array}{cccc}
      0 & 0 & 0 & 0 \\
      0 & v & -v & w \\
      0 & -v & v & -w \\
      0 & w^* & -w^* & q\\
    \end{array}\right).
\end{equation}
Just as in the previous section, we shall consider the case where the four qubits are initially prepared in a state of the form
 $ \ket{\Phi_{A}}\otimes \ket{\Phi_{B}} $ or $\ket{\Psi_{A}}\otimes \ket{\Psi_{B}}$ (see Eqs.(\ref{Phi} \ref{Psi})). For these two cases, tracing out the degrees of freedom corresponding to any pair of qubits we obtain the reduced density matrix for the other pair. Thus, in the long time limit we find that
\begin{equation}\label{A1A2_2}
\rho^{A}=\left(\begin{array}{cccc}
      F_{22} & 0 & 0 & H_{24} \\
      0 & F_{33} & 0 & 0 \\
      0 & 0 & K_{22} & 0 \\
      N_{42} & 0 & 0 & K_{33}+P_{44}\\
    \end{array}\right),
\end{equation}
and
\begin{equation}
\rho^{A_1B_2}=\left(\begin{array}{cccc}
      F_{33} & 0 & 0 & H_{34} \\
      0 & F_{22} & 0 & 0 \\
      0 & 0 & K_{33} & 0 \\
      N_{43} & 0 & O & K_{22}+P_{44}\\
    \end{array}\right).
\end{equation}
In particular, for the case where the qubits are initially in $ \ket{\Phi_{A}}\otimes \ket{\Phi_{B}} $, we
obtain
\begin{equation}
\label{rholong-time1}
\rho^{A}=\left(\begin{array}{cccc}
      y & 0 & 0 & x \\
      0 & y & 0 & 0 \\
      0 & 0 & y & 0 \\
      x^* & 0 & 0 & 1-3y\\
    \end{array}\right),
\end{equation}
and
\begin{equation}
\label{rholong-time2}
\rho^{A_1B_2}=\left(\begin{array}{cccc}
      y & 0 & 0 & -x \\
      0 & y & 0 & 0 \\
      0 & 0 & y & 0 \\
      -x^* & 0 & 0 & 1-3y\\
    \end{array}\right),
\end{equation}
 where $y=1/4\cos^2(\alpha)\sin^2(\alpha)$ and $|x|=1/2\cos(\alpha)\sin^3(\alpha)$. From the above expressions of Eq.~(\ref{rholong-time1}) and Eq.~(\ref{rholong-time2}), we conclude that in the long time limit we have $ C(\rho^{A_1B_2})=C(\rho^{A_1A_2})$.
The concurrence is given by $C=2 \max\{(|x|-y),0\}$. Note that the concurrence does not vanish for partially entangled states having
$\arctan(0.5)\approx26.6^\circ<\alpha\leqslant 90^\circ$. Interestingly, the maximum of the concurrence $C_{max}=0.24$ is attained at
$ \alpha\approx65.3^\circ $ which is consistent with the single mode model discussed in section Sec.\ref{single-mode}.\\

\indent
As for the case ( $\ket{\Psi_{A}}\otimes \ket{\Psi_{B}}$), we find that the long time density matrix now reads
\begin{equation}
\rho^{A}=\left(\begin{array}{cccc}
      y & 0 & 0 & 0 \\
      0 & y & 0 & 0 \\
      0 & 0 & y & 0 \\
      0 & 0 & 0 & 1-3y\\
    \end{array}\right).
\end{equation}
 Therefore there is no entanglement present in the final state of $A_1A_2$. Following the similar steps, $\rho^{A_1B_2}$ also ends up as a separable state for this initial condition.
\section{Conclusions}
In this paper we studied the entanglement dynamics in a generalized double JC model. We showed that although the system evolves non-trivially, two pairs of qubits ($A_{1}A_{2}$ and $B_{1}B_{2}$) can preserve some fraction of their initial entanglement. We found a family of initial states for the system $\ket{\Phi_{A}}\otimes \ket{\Phi_{B}}\otimes \ket{0_{1}}\otimes \ket{0_{2}}$. such that the entanglement in the pair never vanishes. We also determined the optimal initial state for which the concurrence is greater than $0.24$ at all times. Interestingly, this optimal state is not a maximally entangled state.  This result does not involve conditional dynamics (i.e. no quantum measurements are required). The scenario presented in this paper should be compared with the double JC model (see Sec.\ref{Yu-Eberly}) where this preservation is not possible for any initial configuration of the system. Thus, putting aside questions related to the experimental realization of our scenario, the comparison of both models suggest that storing the qubits in pairs may be a way to protect their entanglement. One can envision even larger networks with qubits prepared in multi-particle entangled states. It would be interesting to explore such systems and study the amount of entanglement available at any time. The aforementioned effect of mutual preservation can be interpreted as the result of the constructive interference of the amplitudes corresponding to processes of emission, absorption etc. It may be also interpreted as partial entanglement transfer, that is, the initial entanglement cannot be completely redistributed over the rest of the pairs. In the double JC model the initial entanglement of the pairs can be completely transferred to the cavities \cite{Yonac-Yu}. \\
{\indent}Naturally, one is tempted to study all pairwise quantum correlation and attempt to establish entanglement conservation rules for this model (as in \cite{Sainz-Bjork} and \cite{Ficek}). For mixed states we only know the separability criteria for the low dimensional Hilbert spaces ${\cal C}^M\times{\cal C}^N$ with $M=2$ and $N=2$ or $N=3$ \cite{Hodorecki96} or for the case of bipartite Gaussian states \cite{Simon}. As a result,  all pairwise concurrences can be computed except for $F_{1}F_{2}$ (cavity-cavity) which is, effectively,  a $3\times3 $ system.\\
 Finally in Sec.\ref{multi-mode} we included multimode qubit-cavity interactions and studied the dynamics of the system by means of a non-Markovian state diffusion equation. We found the density matrices in the long time limit. The results corroborate those from the single-mode interaction model.\\
{\indent} The latter suggests that it would be interesting to explore other multi-qubit configurations and interaction models. Such studies may lead to a better understanding of the entanglement dynamics and provide interesting insights into the problem of protecting entanglement from the environment.
\section*{Acknowledgments}
A.V. acknowledges the support of the Intelligence Community Postdoctoral Research Fellowship. A.V. and C. W.W. acknowledge support from DARPA (W911NF-10-1-0416), NSF CAREER award (ECCS-0747787), and NSF IGERT (DGE-1069240), J.J and T.Y.  acknowledge support from NSF PHY- 0925174, and DOD/AF/AFOSR No. FA9550-12-1-0001.
\section{Appendix}
\label{appendix}
The operators $\mV_{ijkl}$ operators defined in Eq.(\ref{main2}) satisfy the properties $\mV_{ijkl}^{\dagger}=\mV_{lkji}$ and $\sum_{k}\mV_{ikkl}=\delta_{il}\mathbb{I}$. When the cavities are prepared in the pure state $\rho_{F_{i}}=\ket{N}\bra{N}$ these assume the form
\begin{eqnarray}
\mV_{1111}&=&\braket{N|\braket{e|{\mU_{i}}^{\dagger}|e}\braket{e|\mU_{i}|e}|N}= \left(
                                                                             \begin{array}{cc}
                                                                               a & 0 \\
                                                                               0 & b \\
                                                                             \end{array}
                                                                           \right)\\
\mV_{1121}&=&\braket{N|\braket{e|{\mU_{i}}^{\dagger}|e}\braket{g|\mU_{i}|e}|N}= \left(
                                                                             \begin{array}{cc}
                                                                               0 & c \\
                                                                               0 & 0 \\
                                                                             \end{array}
                                                                           \right) e^{i \gamma} \\
\mV_{1221}&=&\braket{N|\braket{e|{\mU_{i}}^{\dagger}|g}\braket{g|\mU_{i}|e}|N}= \left(\begin{array}{cc}
                                                                             d & 0 \\
                                                                             0 & f \\
                                                                           \end{array}  \right)\\
\mV_{2112}&=&\braket{N|\braket{g|{\mU_{i}}^{\dagger}|e}\braket{e|\mU_{i}|g}|N}= \left(\begin{array}{cc}
                                                                             h & 0 \\
                                                                             0 & k \\
                                                                           \end{array}  \right)\\
\mV_{2122}&=& \braket{N|\braket{g|{\mU_{i}}^{\dagger}|e}\braket{g|\mU_{i}|g}|N}= \left(\begin{array}{cc}
                                                                             0 & l \\
                                                                             0 & 0\\
                                                                           \end{array}  \right)e^{i\gamma}\\
\mV_{2222}&=&  \braket{N|\braket{g|{\mU_{i}}^{\dagger}|g}\braket{g|\mU_{i}|g}|N}= \left(
                                                                               \begin{array}{cc}
                                                                                 m & 0 \\
                                                                                 0 & n \\
                                                                               \end{array}
                                                                             \right)\\
\mV_{1112}&=& \braket{N|\braket{e|{\mU_{i}}^{\dagger}|e}\braket{e|\mU_{i}|g}|N}=\left(
                                                                          \begin{array}{cc}
                                                                            0 & 0 \\
                                                                            p & 0 \\
                                                                          \end{array}
                                                                        \right)\\
\mV_{1122}&=& \braket{N|\braket{e|{\mU_{i}}^{\dagger}|e}\braket{g|\mU_{i}|g}|N}= \left(
                                                                              \begin{array}{cc}
                                                                                q & 0 \\
                                                                                0 & r \\
                                                                              \end{array}
                                                                            \right)e^{i \gamma}\\
 \mV_{1222}&=& \braket{N|\braket{e|{\mU_{i}}^{\dagger}|g}\braket{g|\mU_{i}|g}|N}= \left(
                                                                          \begin{array}{cc}
                                                                            0 & 0 \\
                                                                            -p & 0 \\
                                                                          \end{array}
                                                                        \right)
                                                                        \end{eqnarray}
                                                                        where $\gamma=\omega t$. Additionally we have
$\mV_{1211}=\mV_{1121}^{\dagger}$, $\mV_{2212}=\mV_{2122}^{\dagger}$, $\mV_{2211}=\mV_{1122}^{\dagger}$, $\mV_{2111}=\mV_{1112}^{\dagger}$, $\mV_{2221}=\mV_{1222}^{\dagger}$
and $\mV_{1212}=\mV_{2121}=0$, which completes the list. The functions $a,b,c \ldots $  read

\begin{eqnarray}
a&=&(1-\frac{N+1}{N+3/2}\sin^{2}(\lt \sqrt{N+3/2}))^2 \nonumber \\
&+&\frac{N+1}{4(N+3/2)}\sin^{2}(2\lt\sqrt{N+3/2})\\
b&=&\cos^{4}(\lt\sqrt{N+1/2})\nonumber \\
&+&\frac{N}{4(N+1/2)}\sin^{2}(2\lt\sqrt{N+1/2})\\
c&=&\frac{N+1}{4 \sqrt{(N+1)^{2}-1/4}}\sin(2\lt\sqrt{N+1/2})\sin(2\lt\sqrt{N+3/2}) \nonumber \\
                                  &-&\sin^{2}(\lt\sqrt{N+1/2})(1-\frac{N+1}{N+3/2}\sin^{2}(\lt\sqrt{N+3/2}))\\
d&=&\frac{N+1}{4(N+3/2)}\sin^{2}(2\lt\sqrt{N+3/2})\nonumber \\
&+&\frac{(N+1)(N+2)}{(N+3/2)^{2}}\sin^{4}(\lt\sqrt{N+3/2})\\
f&=& \sin^{4}(\lt\sqrt{N+1/2})+\frac{N+1}{4(N+1/2)}\sin^{2}(2\lt\sqrt{N+1/2})\\
h&=&\sin^{4}(\lt\sqrt{N+1/2})+\frac{N}{4(N+1/2)}\sin^{2}(2\lt \sqrt{N+1/2})\\
k&=&\frac{N}{4 (N-1/2)}\sin^{2}(2\lt\sqrt{N-1/2})\nonumber \\
&+&\frac{N(N-1)}{(N-1/2)^{2}}\sin^{4}(\lt\sqrt{N-1/2})\\
l&=&\frac{N}{4(\sqrt{N^{2}-1/4})}\sin(2\lt\sqrt{N-1/2})\sin(2\lt\sqrt{N+1/2})\nonumber \\
&-&\sin^{2}(\lt\sqrt{N+1/2})(1-\frac{N}{N-1/2}\sin^{2}(\lt\sqrt{N-1/2}))\\
m&=&\cos^{4}(\lt\sqrt{N+1/2})\nonumber\\
&+&\frac{N+1}{4(N+1/2)}\sin^{2}(2\lt\sqrt{N+1/2})\\
n&=&\frac{N}{4(N-1/2)}\sin^{2}(2\lt\sqrt{N-1/2})\nonumber \\
&+&(1-\frac{N}{N-1/2}\sin^{2}(\lt\sqrt{N-1/2}))^2\\
p&=& -\frac{1}{8(N+1/2)}\sin^{2}(2\lt\sqrt{N+1/2})\\
q&=& \cos^{2}(\lt \sqrt{N+1/2})(1-\frac{N+1}{N+3/2}\sin^{2}(\lt\sqrt{N+3/2}))\nonumber \\
&+&\frac{N+1}{4\sqrt{(N+1)^{2}-1/4}}\sin(2\lt\sqrt{N+1/2}) \nonumber \\
&\times& \sin(2\lt\sqrt{N+3/2})\\
r&=&\cos^{2}(\lt \sqrt{N+1/2})(1-\frac{N}{N-1/2}\sin^{2}(\lt \sqrt{N-1/2}))\nonumber \\
&+&\frac{N}{4 \sqrt{N^{2}-1/4}}\sin(2\lt\sqrt{N-1/2})\sin(2\lt \sqrt{N+1/2})\\ \nonumber
\end{eqnarray}

 \end{document}